\documentclass[journal=jpclcd,manuscript=letter,layout=traditional]{achemso}

\usepackage[version=3]{mhchem} 
\usepackage[T1]{fontenc}       
\usepackage{setspace}  
\usepackage[font={small,sf,md}]{caption}  

\author{Phil Rosenow}
\affiliation{Fachbereich Chemie, Philipps-Universit\"at Marburg, Hans-Meerwein-Stra\ss e 4, 35032 Marburg, Germany}
\author{Peter Jakob}
\affiliation{Fachbereich Physik, Philipps-Universit\"at Marburg, Renthof 5, 35032 Marburg, Germany}
\alsoaffiliation{Material Sciences Center, Philipps-Universit\"at Marburg, Hans-Meerwein-Stra\ss e, 35032 Marburg, Germany}
\author{Ralf Tonner}
\affiliation{Fachbereich Chemie, Philipps-Universit\"at Marburg, Hans-Meerwein-Stra\ss e 4, 35032 Marburg, Germany}
\alsoaffiliation{Material Sciences Center, Philipps-Universit\"at Marburg, Hans-Meerwein-Stra\ss e, 35032 Marburg, Germany}
\email{tonner@chemie.uni-marburg.de}
\phone{+49 (0)6421 2825418}
\fax{+49 (0)6421 2821826}

\title[]
  {Electron-Vibron Coupling at Metal-Organic Interfaces from Theory and Experiment}

\begin{document}
\sffamily  

\begin{tocentry}
\includegraphics[width=5cm]{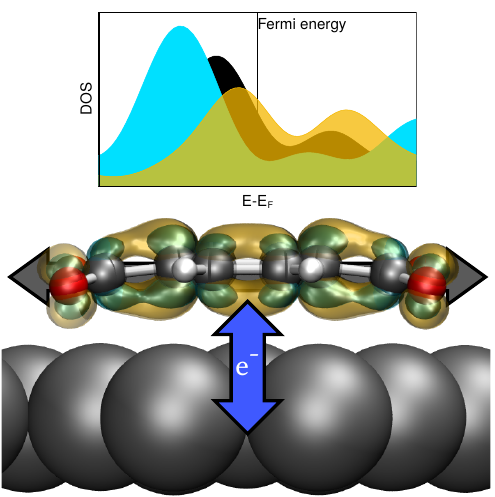}
\end{tocentry}

\begin{abstract}
We study the significance and characteristics of interfacial dynamical charge transfer at metal-organic interfaces for the organic semiconductor model system 1,4,5,8-naphthalene\-tetra\-carboxylic dianhydride (NTCDA) on Ag(111) quantitatively. We combine infrared absorption spectroscopy and dispersion-corrected density functional theory calculations to analyze dynamic dipole moments  and electron-vibron coupling at the interface. We demonstrate that interfacial dynamical charge transfer is the dominant cause of infrared activity in these systems and that it correlates with results from partial charge and density of states analysis. Nuclear motion generates an additional dynamic dipole moment but represents a minor effect except for modes with significant out-of-plane amplitudes.
\end{abstract}

\normalsize  
\onehalfspacing  

Metal-organic interfaces are essential elements influencing the performance of molecular electronics applications like light-emitting devices or field-effect transistors. 
Here, electronic excitations via charge-transfer excitons often play a key role.
The formation mechanisms of these excitons via electron-hole pair (EHP) formation at the interface thus need to be understood. The coupling of EHP to adsorbate vibrations (electron-vibron coupling) is one way to induce an interfacial dynamical charge transfer (IDCT), which delivers key information about the electronic structure at the interface.
IDCT induced by an orbital dipping in and out of the Fermi sea has been suggested for small molecules on metal surfaces \cite{Persson1980, Persson1987, Persson1990, Zhdanov1988,Arnolds2011}. This model has been applied to larger organic adsorbates on metal surfaces more recently \cite{Tautz2002, Tautz2002b, Tautz2007, Eremtchenko2003}. The effect of such coupling on the vibrational line shape was analyzed by Langreth \cite{Langreth1985}. Despite its importance, there is no theoretical or experimental proof for IDCT as such and the relative importance compared to nuclear motion at a metal-organic interface beyond heuristic models up to now. We quantify this contribution by first-principles analysis and provide a rationale for the amount of IDCT for specific vibrational modes at the interface.

Prerequisites for the occurrence  of IDCT are (i) partial occupation of an adsorbate molecular orbital upon adsorption and (ii) strong electron-vibron coupling for relevant substrate electronic levels \cite{Chabal1985}. This was met in early studies for CO on Cu(100) \cite{Persson1980} and O$_2$ on Pt(111) \cite{Persson1987}.
Later, IDCT was observed in studies on fullerene $\mathrm{C_{60}}$ \cite{Peremans1997,Silien1999,Rudolf2002}.
IDCT, in conjunction with EHP excitation is also held responsible for the strong  damping of high frequency vibrations of adsorbates via energy dissipation, which was one of the driving forces for its analysis \cite{Persson1980, Krishna2006, Langreth1985}. Line-width analysis for IR\ measurements then leads to an estimate of vibrational lifetimes \cite{Wodtke2008}.
The occurrence of electron-vibron coupling is essentially a breakdown of the Born-Oppenheimer approximation. For systems with strong non-adiabaticity, IDCT thus leads to asymmetric line shapes in vibrational spectra \cite{Langreth1985,Chabal1985}. While a theoretical treatment of this asymmetry  requires  a time-dependent analysis \cite{Saalfrank2011}, which is not feasible for the system investigated here, an empirical evaluation for NTCDA on Ag(111) has been presented recently by Braatz \latin{et al.} \cite{Braatz2012}.

In this Letter, we will present a microscopic view on the vibrational modes of 1,4,5,8-naphthalene\-tetra\-carboxylic dianhydride (NTCDA) on Ag(111) based on experiment and computation focusing on the contribution of IDCT to the observed infrared (IR) absorption intensities. We correlate this to results of charge-transfer analysis associated with the electron-vibron coupling. We demonstrate that IDCT is the dominating effect for electron-vibron coupling in this system, quantify the charge-transfer associated and verify that IDCT is crucial for symmetric $\mathrm{a_g}$ modes to become IR-active, in stark contrast to the free molecule.
The system chosen here constitutes a model system for a more general class of metal-organic interfaces with noble metal substrates and planar $\pi$-conjugated organic semiconductor species and we expect the results to be transferable to systems with similar energy level matching.
Quantification of electron-vibron coupling has been identified as a major challenge for these interfaces \cite{Draxl2014}.

Experimentally, infrared absorption spectroscopy (IRAS) is the method of choice to extract information regarding IDCT since it provides accurate vibrational energy and intensity measurements for modes originating at the interface. Furthermore, IRAS adheres to strict selection rules and probes molecular layers in a non-destructive way. 
The IR intensity of a mode is proportional to the square of the change of dipole moment of this vibration with respect to the equilibrium structure (dynamic dipole moment) 
\begin{equation}
 I_{IR} \propto \mu_{dyn} ^2,
\label{eq:IR}
\end{equation}
where $\mu_{\mathrm{dyn}}$ results from contributions of nuclear motion ($\mu_{\mathrm{nucl}}$) and dynamic charge transfer across the adsorbate-substrate interface ($\mu_{\mathrm{IDCT}}$)
\begin{equation}
	\mu_{dyn} = \mu_{IDCT} + \mu_{nucl},
\label{eq:dyn}
\end{equation}
as depicted in Scheme \ref{fig:my-dyn}c.
Upon adsorption, charge q is being transferred from the surface to the adsorbate. Changes in q along vibrational coordinate Q (dynamic charge transfer $\Delta$q) can be determined by partial charge analysis; they are proportional to the IDCT contribution:
 \begin{equation}
	\mu_{IDCT} \propto \Delta \text{q}.
\label{eq:IDCT}
\end{equation}

We will show that the contribution $\mu_{\mathrm{nucl}}$ can be derived from equation \ref{eq:dyn}, if $\mu_{\mathrm{IDCT}}$ is the major component. Additionally, the contribution from nuclear motion can be estimated directly from 
 \begin{equation}
	\mu^{'}_{nucl} \propto \mu_{\Delta\text{Q}},
\label{eq:nucl}
\end{equation}
where $\mu_{\Delta\text{Q}}$ gives the dipole moment change of the adsorbate distorted along Q without the presence of the surface.  
For an adsorbate on a metal substrate only the z-component of $\mu_{dyn}$ contributes to the intensity due to dipole selection rules.
Therefore, $\mu_{\mathrm{	nucl}}$ is determined by out-of-plane displacements of the adsorbate atoms (Scheme \ref{fig:my-dyn}c).

\begin{scheme}
\centering
\includegraphics[width=8cm]{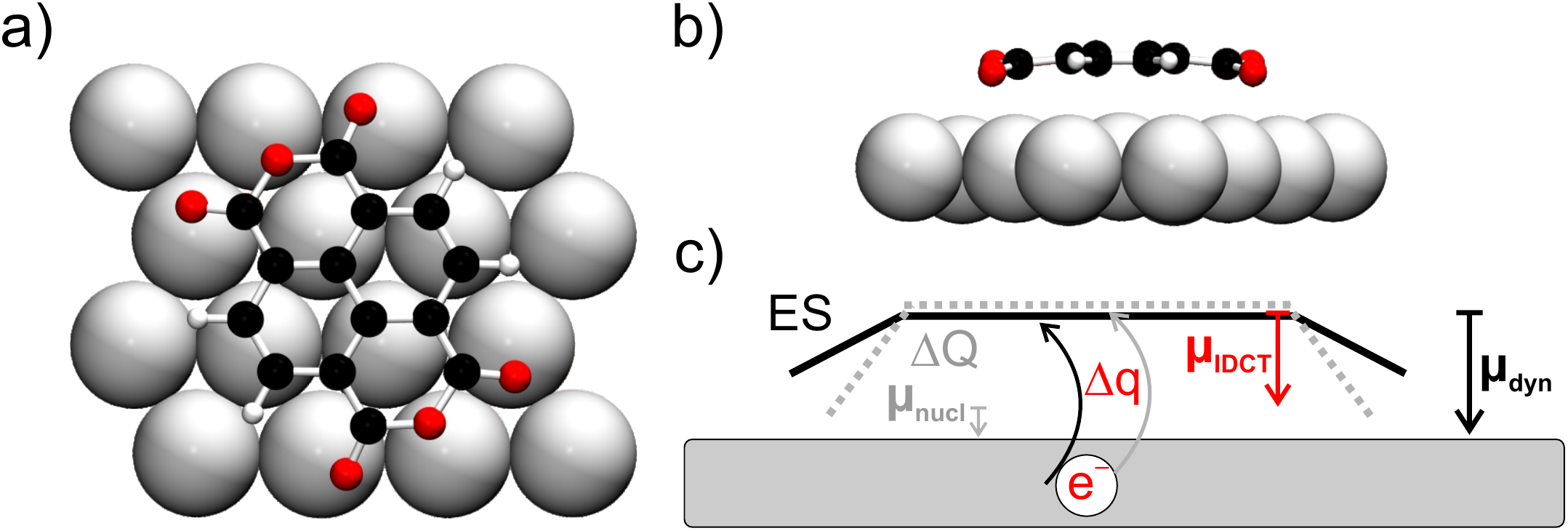}
\caption{Most stable adsorption geometry of NTCDA on a bridge position on Ag(111) in top (a) and side (b) view; (c) contributions to dynamic dipole moment.}
\label{fig:my-dyn}
\end{scheme}

Atomic and electronic ground state structures of metal-organic interfaces are well represented by dispersion-corrected density functional theory (DFT) \cite{Romaner2009, Bauer2012}. The description of vibrational spectra has also successfully been achieved in the past \cite{Baroni2001,Breuer2012}.

NTCDA represents a model system with typical properties for an important class of organic semiconductors suitable for metal-organic interfaces: (i) Highly symmetric planar atomic structure, (ii) delocalized $\pi$-electron system, (iii) strong binding to metal surfaces with multiple bonding mechanisms. The totally symmetric $\mathrm{a_g}$ modes are IR-inactive for the free molecule ($\mu_{\mathrm{dyn}} = 0$). According to IRAS the vibrational spectrum of NTCDA on Ag(111) is strongly dominated by these in-plane vibrational modes, which suggests a notable contribution to $\mu_{\mathrm{dyn}}$ from IDCT \cite{Braatz2012}. Near edge x-ray absorption fine structure measurements showed strong electron-vibron coupling in this system which renders it highly suitable for the quantitative determination of IDCT \cite{Schoell2004}.

Our computational investigations applied dispersion-corrected DFT computations at the GGA level (PBE-D3) \cite{PBE,GrimmeD3,GrimmeD3BJ} as outlined in the computational details section. The sub-monolayer structure investigated here differs from a genuine relaxed monolayer insofar as the NTCDA molecules are separated by one extra Ag atom row so that hydrogen bonding between neighbouring adsorbates can be neglected and all NTCDA occupy identical adsorption sites.  NTCDA preferentially adsorbs with the central C=C bond occupying a bridge position on the Ag(111) surface as shown in Scheme~\ref{fig:my-dyn}a. The detailed discussion of structural parameters, which agree well with experimental data of x-ray standing wave measurements,\cite{Stadler2007} is presented elsewhere \cite{Tonner2015_NTCDA_fullpaper}.
The molecule is bonded by van-der-Waals interactions together with direct oxygen-silver chemical bonds (Scheme \ref{fig:my-dyn}b) leading to a strong adsorption and thus fulfilling the first prerequisite for IDCT \cite{Tautz2007}. The partial filling of the LUMO upon adsorption is a well-known phenomenon for NTCDA \cite{Bendounan2007} and thus the second requirement for IDCT is also satisfied. 

\begin{figure} 
\centering
\includegraphics[width=8cm]{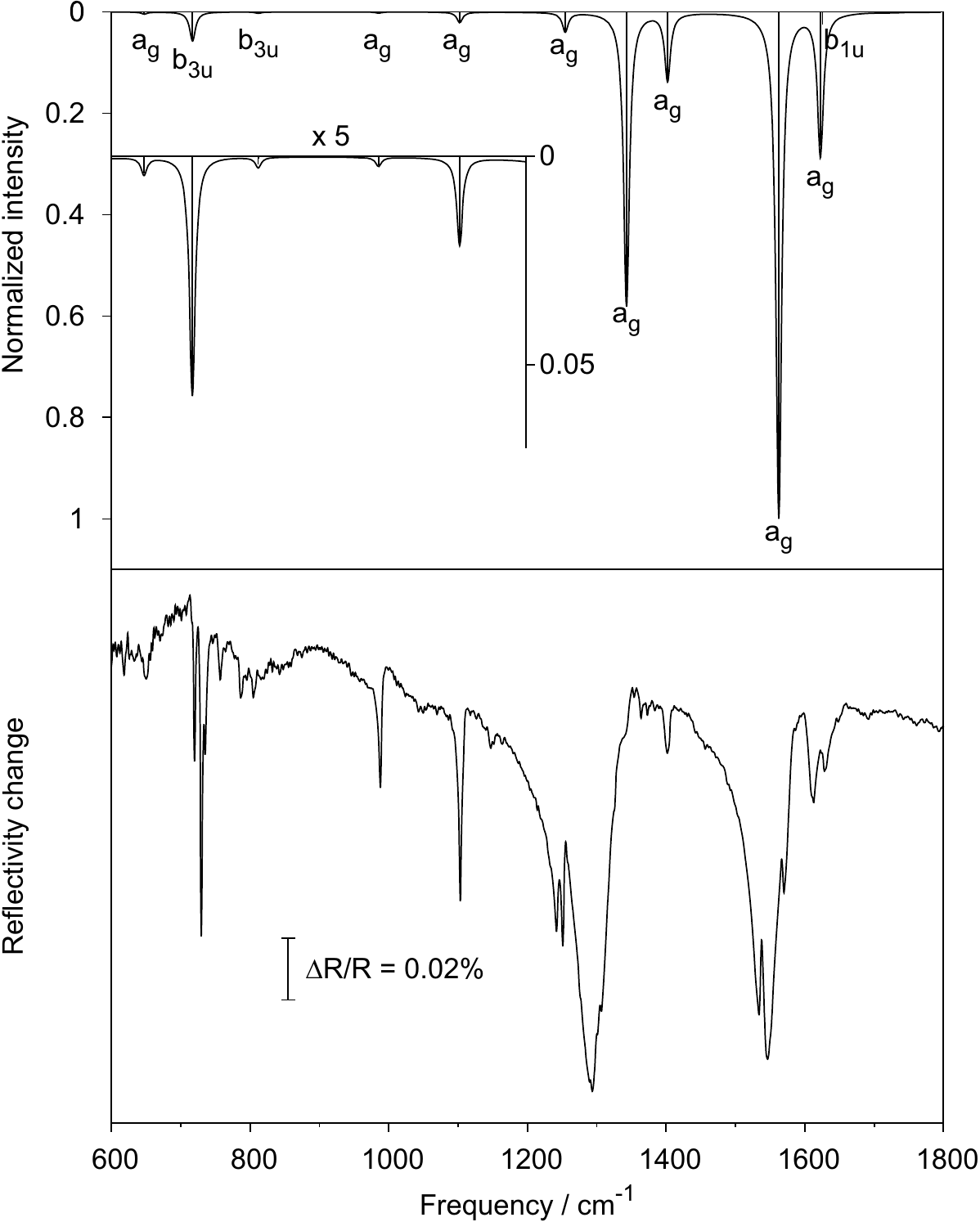}
\caption{Computed (top) and measured (bottom) IR spectrum for NTCDA on Ag(111). Intensity scaling (x5, inset)\ was applied to signals below $1200\;\mathrm{cm}^{-1}$ in the computed spectrum. Flags indicate symmetry of respective modes applying free molecule nomenclature.}
\label{fig:spectrum}
\end{figure}

In Figure~\ref{fig:spectrum} the experimental and computed IR spectra of NTCDA on Ag(111) are displayed. 
The IRAS spectrum refers to 0.15 monolayers deposited at 28 K. The low temperature ensures that isolated NTCDA molecules are present and the formation of denser islands is avoided, in accordance with the arrangement used in our theoretical analysis. Comparison to  previously derived spectra for the relaxed monolayer of NTCDA on Ag(111)\cite{Braatz2012}  yield moderate frequency shifts and only slight intensity variations, most probably due to some minor structural modification as a result of the different local environment. 
We obtain very good agreement of experiment and theory for vibrational energies and relative intensities. Thus, we can trust the computational spectrum to reproduce all the essential features and base the further examination on these data.

The determination of mode symmetry for the vibrations of the adsorbate has been achieved by comparison of the displacement patterns to the free NTCDA molecule. Although the molecule bends upon adsorption (Scheme \ref{fig:my-dyn}b) and the point group symmetry reduces from $\mathrm{D_{2h}}$ to approximately $\mathrm{C_{2v}}$, the respective symmetry labels of modes are maintained as this way of labeling is the most helpful and intuitive one. In Figure \ref{fig:spectrum} it is apparent that most of the intense IR bands belong to the totally symmetric $\mathrm{a_g}$ irreducible representation. Notably, these modes are IR-inactive in the free molecule since they do not exhibit dynamic dipole moments.

\begin{figure} 
\centering
\includegraphics[width=8cm]{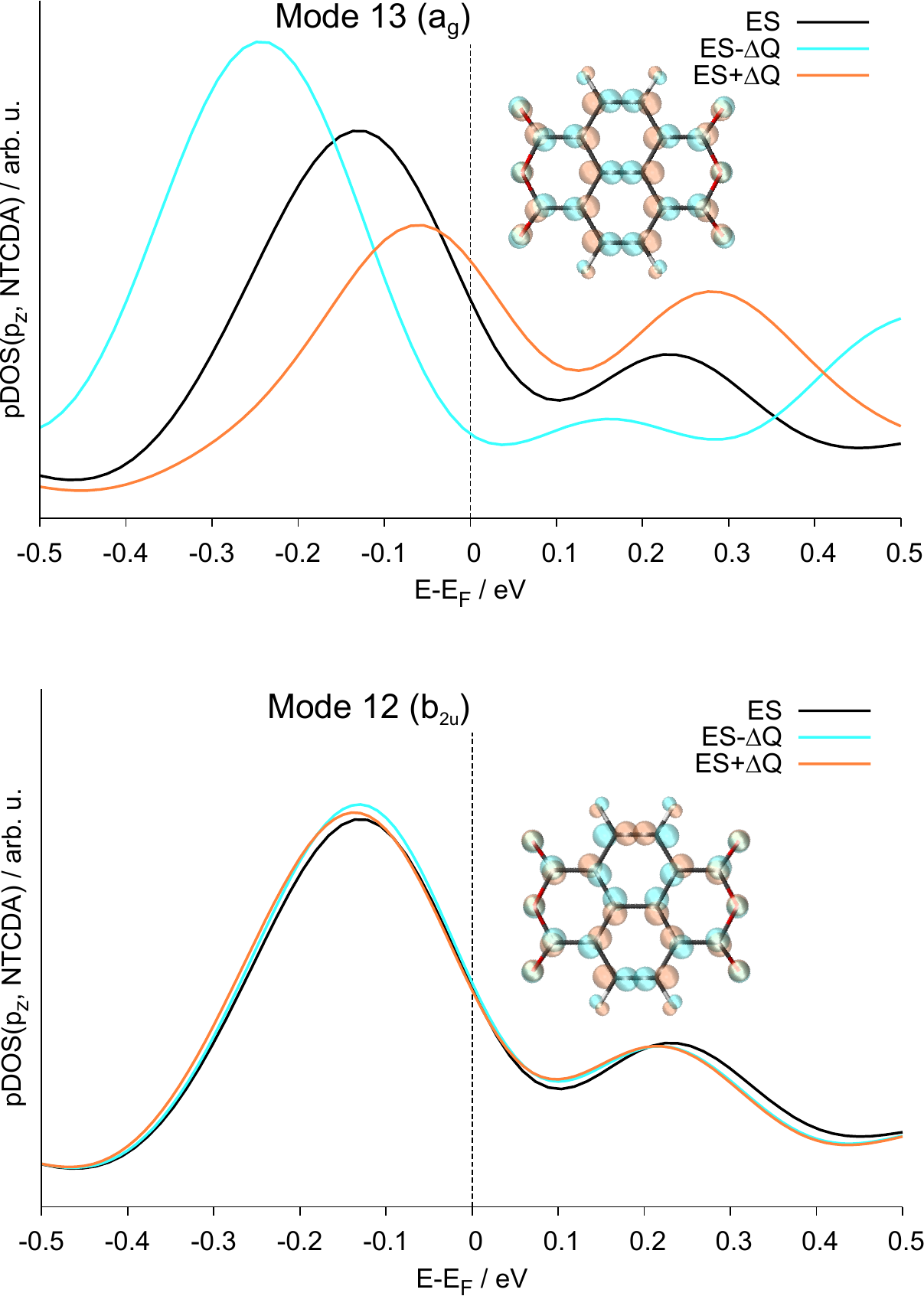}
\caption{Equilibrium Structure (ES) of NTCDA with positive (ES+$\Delta$Q, orange) and negative (ES-$\Delta$Q, light blue) distortion along mode 13 ($1565.6\;\mathrm{cm^{-1}}$, $\mathrm{a_g}$, top) and mode 12 ($1509.8\;\mathrm{cm^{-1}}$, $\mathrm{b_{2u}}$, bottom) together with the corresponding pDOS stemming from $p_z$ states of NTCDA. The atomic displacements have been amplified.}
\label{fig:disps}
\end{figure}

Next, we analyze the contribution of IDCT to $\mathrm{I_{IR}}$ by displacing the adsorbate along normal mode coordinate Q. For the most intense mode at $1565.6\;\mathrm{cm}^{-1}$ ($\mathrm{a_g}$) the displacement pattern and the corresponding pDOS of the $\pi$-orbitals (composed of $\mathrm{p_z}$ atomic orbitals) of NTCDA are displayed in Figure~\ref{fig:disps} (top). Shown is the pDOS for the equilibrium structure (black line) together with positively (orange line) and negatively (blue line) displaced structures. The partial occupation of the bands stemming from $\pi$-orbitals is unveiled (Fermi level crossing the pDOS lines), thus reflecting the known partial LUMO occupation upon adsorption. For the displaced structures, we see a strong shift of the pDOS maximum and a respective change in density at the Fermi level. Thereby, we  observe electron-vibron coupling in this system and identify a strong dependence of the electronic structure on the vibrational motion -- a direct prove for IDCT derived from first principles calculations rather than heuristic assumptions. Analogous behavior is observed for the other $\mathrm{a_g}$ modes in the adsorbate, albeit at a reduced extent.
Analyzing a larger entity of modes (Table \ref{tab:dispdata}), we find that IR-active modes exhibit notable values for $\mu_{dyn}$ as expected from equation \ref{eq:IR}. The high values for $\Delta$q for most IR-active modes are a strong indication that IDCT is the leading effect in the system investigated.

Further support for this hypothesis can be found by analyzing the modes exhibiting different symmetries. Modes 3 ($717.9\;\mathrm{cm}^{-1}$, $\mathrm{b_{3u}}$), 5 ($813.0\;\mathrm{cm}^{-1}$, $\mathrm{b_{3u}}$) and 15 ($1628.9\;\mathrm{cm}^{-1}$, $\mathrm{b_{1u}}$) exhibit significant, albeit much lower intensities in the experimental as well as the computed spectrum (Figure ~\ref{fig:spectrum}, Table \ref{tab:dispdata}) in line with their lower $\Delta$q values.
For the $\mathrm{b_{3u}}$-symmetric modes 3 and 5, we find a significant out-of-plane bending component as expected from symmetry considerations.
This leads to a major contribution to $\mu_{dyn}$ that can be quantified by determining $\mu_{\Delta\text{Q}}$ (Scheme \ref{fig:my-dyn}), which shows that mode 3 has the highest contribution from nuclear motion \bibnote{Dynamic dipole moments for adsorbate-only calculations were determined by PBE/def2-TZVPP computations with the code Gaussian09. $\mu_{\Delta\text{Q}}$ were then derived by adding up the dipole moment changes from calculations of distortions in $\mathrm{+} \Delta\text{Q}$ and $\mathrm{-}\Delta\text{Q}$ directions.}. 

\begin{figure} 
\centering
\includegraphics[width=16cm]{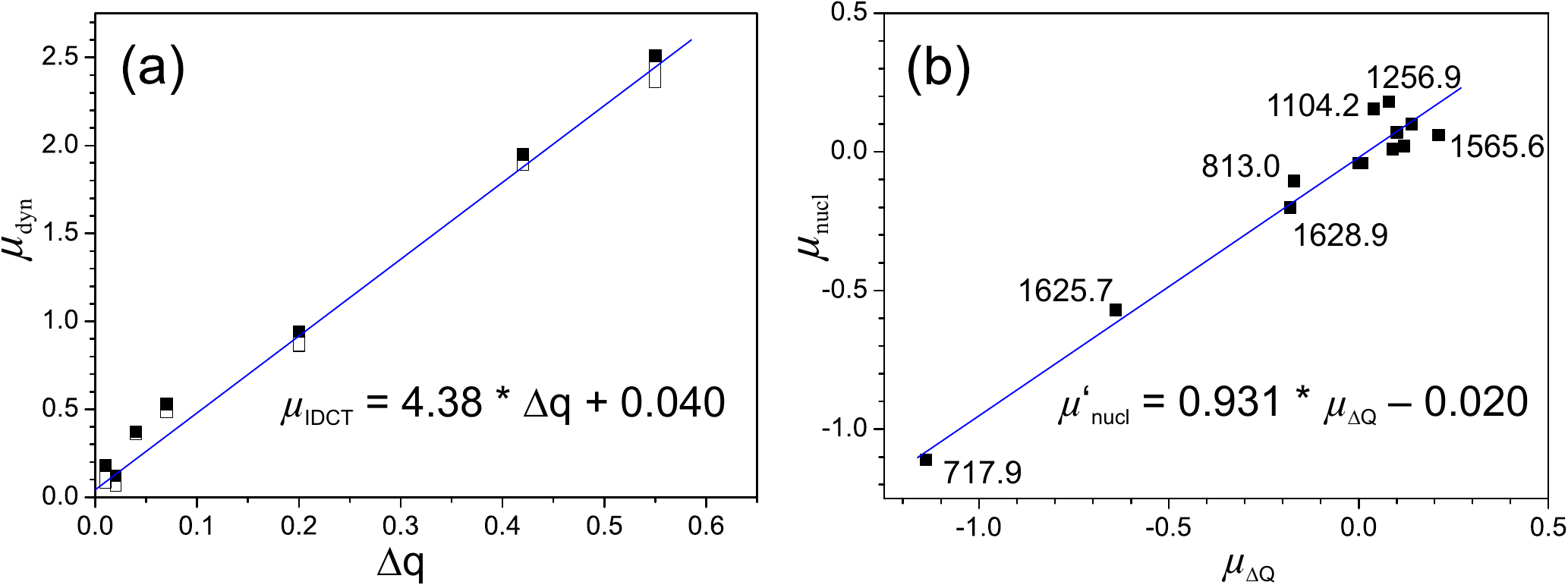}
\caption{(a) Correlation of charge transfer ($\mathrm{\Delta\text{q}}$) and $\mathrm{\mu_{dyn}}$ to determine $\mathrm{\mu_{IDCT}}$; the open rectangles refer to first order correction of $\mathrm{\mu_{dyn}}$ due to nonzero $\mathrm{\mu_{nucl}}$. (b) Correlation of $\mathrm{\mu_{\Delta\text{Q}}}$ with $\mathrm{\mu_{nucl}}$ to determine $\mathrm{\mu^{'}_{nucl}}$.} 
\label{fig:corr}
\end{figure}

Although we now established qualitative relations between $\mu_{dyn}$ and its contributions from IDCT and nuclear motion, we aim at a direct relationship to allow the quantitative comparison of both factors. Figure \ref{fig:corr}a reveals that the charge transfer contribution $\Delta$q indeed strongly correlates with $\mu_{dyn}$. No correlation is found with $\mu_{\Delta\text{Q}}$. Although correlation does not imply causality, combined with the pDOS analysis it becomes clear that IDCT determines $\mu_{dyn}$ of $\mathrm{a_g}$ type of modes and thereby $\mathrm{I_{IR}}$. $\mu_{IDCT}$ can now be derived from $\Delta$q by the linear correlation given in Figure \ref{fig:corr}a under the assumption that it is the leading term ($\mu_{IDCT} \approx \mu_{dyn}$) \bibnote{Modes without significant $\mu_{nucl}$ were considered in the fit.} and is given in Table \ref{tab:dispdata}. From equation \ref{eq:dyn}, the signed value for $\mu_{nucl}$ can then readily be determined as the difference of $\mu_{dyn}$ and $\mu_{IDCT}$. The sign convention is thus that a negative value indicates $\mu_{nucl}$ pointing in the opposite direction compared to $\mu_{IDCT}$. The good correlation of $\mu_{nucl}$ with the computed $\mu_{\Delta\text{Q}}$ in Figure \ref{fig:corr}b validates the approach. This linear correlation can now be employed to convert $\mu_{\Delta\text{Q}}$ values into $\mu^{'}_{nucl}$ (eq. \ref{eq:nucl}) given in Table \ref{tab:dispdata}. We can thus derive $\mu^{'}_{nucl}$ directly from $\mu_{\Delta\text{Q}}$ or as difference from $\mu_{IDCT}$, while the latter approach gives information regarding the relative sign of the contributions. 

For most modes investigated, we find $\mu_{IDCT}$ to be the leading contribution to $\mu_{dyn}$ (Table \ref{tab:dispdata}). More specifically, for three out of the four modes with highest intensity (9, 10, 13), IDCT effects are an order of magnitude larger. The third-most intense mode 14 ($1625.7\;\mathrm{cm}^{-1}$) also shows high IDCT contributions but lower $\mathrm{I_{IR}}$. This can be understood by the opposite sign of the strong $\mu_{nucl}$ contribution, which stems from a substantial vibrational amplitude for the downward-bent acyl-oxygen atoms at the periphery of the adsorbate. Moreover, analysis of the displacement patterns unveils a mixed nature for modes 14 and 15 
with dominant $\mathrm{a_g}$ ($\mathrm{b_{1u}}$) contributions. We suspect that this mixing of character is provoked by their close spectral vicinity. This also explains the surprising intensity of mode 15 since $\mathrm{b_{1u}}$ symmetry should render it dipole forbidden on a metal surface ($\mu_{\mathrm{dyn}}$ along NTCDA long axis). This example shows the limit of the assignments based on molecular mode symmetries which is clear-cut for all other vibrations and the need to study the contributions to $\mu_{\mathrm{dyn}}$ in detail.
Modes with medium intensity (6, 7, 8) and $\mathrm{a_g}$ symmetry exhibit higher relative contributions from nuclear motion but in all cases IDCT is the leading term here as well, while mode 1 shows equivalent contributions from both mechanisms. Modes 3 and 5 ($\mathrm{b_{3u}}$ symmetry), exhibit opposite signs for both terms: This leads to a near-zero intensity for mode 5 ($\mu_{nucl} \approx \mu_{IDCT}$) and high intensity for mode 3 due to nuclear motion ($\mu_{nucl} > \mu_{IDCT}$).

\begin{table*}
\centering
\caption{Computed properties of vibrational modes. See Scheme \ref{fig:my-dyn} for definition of terms.}
\label{tab:dispdata}
\begin{tabular}{rrlcrcrcr}
\hline
No. & \multicolumn{1}{c}{$\tilde{\nu}$\textsuperscript{\emph{a}}} & \multicolumn{1}{c}{sym.\textsuperscript{\emph{b}}} & \multicolumn{1}{c}{int.\textsuperscript{\emph{c}}} & \multicolumn{1}{c}{$\mu_{\mathrm{dyn}}$\textsuperscript{\emph{d}}} & \multicolumn{1}{c}{$\mu_{\mathrm{IDCT}}$\textsuperscript{\emph{d}}} & \multicolumn{1}{c}{$\mu^{'}_{\mathrm{nucl}}$\textsuperscript{\emph{d}}} & \multicolumn{1}{c}{$\Delta$q\textsuperscript{\emph{e}}} & \multicolumn{1}{c}{$\mu_{\mathrm{\Delta\text{Q}}}$\textsuperscript{\emph{d}}}  \\ 
\hline
 1 & 648.1  & $\mathrm{a_g}$    & 0.004 & 0.18 & 0.08 & +0.11 & 0.01 & +0.14 \\
 2 & 665.4  & $\mathrm{b_{2g}}$ & 0.000 & 0.00 & -    &  -    & 0.00 &  0.00 \\
 3 & 717.9  & $\mathrm{b_{3u}}$ & 0.058 & -0.59& 0.52 & -1.08 & 0.11 & -1.14 \\
 4 & 720.8  & $\mathrm{a_u}$    & 0.000 & 0.00 & -    &  -    & 0.00 & +0.01 \\
 5 & 813.0  & $\mathrm{b_{3u}}$ & 0.002 & 0.11 & 0.22 & -0.18 & 0.04 & -0.17 \\
 6 & 987.0  & $\mathrm{a_g}$    & 0.002 & 0.12 & 0.13 & +0.06 & 0.02 & +0.09 \\
 7 & 1104.2 & $\mathrm{a_g}$    & 0.021 & 0.37 & 0.22 & +0.02 & 0.04 & +0.04 \\
 8 & 1256.9 & $\mathrm{a_g}$    & 0.038 & 0.53 & 0.35 & +0.05 & 0.07 & +0.08 \\
 9 & 1345.5 & $\mathrm{a_g}$    & 0.581 & 1.95 & 1.88 & +0.07 & 0.42 & +0.10 \\
10 & 1404.8 & $\mathrm{a_g}$    & 0.133 & 0.94 & 0.92 & +0.09 & 0.20 & +0.12 \\
11 & 1435.8 & $\mathrm{b_{1u}}$ & 0.000 & 0.02 & -    &  -    & 0.00 &  0.00 \\
12 & 1509.8 & $\mathrm{b_{2u}}$ & 0.000 & 0.00 & -    &  -    & 0.00 &  0.00 \\
13 & 1565.6 & $\mathrm{a_g}$    & 1.000 & 2.51 & 2.44 & +0.17 & 0.55 & +0.21 \\
14 & 1625.7 & $\mathrm{a_g}$    & 0.264 & 1.27 & 1.84 & -0.62 & 0.41 & -0.64 \\
15 & 1628.9 & $\mathrm{b_{1u}}$ & 0.024 & 0.37 & 0.57 & -0.19 & 0.12 & -0.18 \\
\hline
\end{tabular}

\textsuperscript{\emph{a} Vibrational modes in $\mathrm{cm}^{-1}$.}
\textsuperscript{\emph{b} Mode symmetries.}
\textsuperscript{\emph{c} IR intensities normalized to highest value.}
\textsuperscript{\emph{d} Dipole moments in Debye.}
\textsuperscript{\emph{e} Charges in e.}
\end{table*}

For modes without significant intensities (i.e. 2, 4, 11, 12 in Table \ref{tab:dispdata}), none of the two mechanisms is active.
Exemplary, for the $\mathrm{b_{2u}}$-symmetric mode at $1509.8\;\mathrm{cm^{-1}}$, the pDOS is represented in Figure~\ref{fig:disps} (bottom). No dynamical change of the electron density at the Fermi level is found in line with negligible contributions from $\mu_{IDCT}$. Thus, although several of these vibrations are IR-active in the free molecule they do not show any intensity on the Ag(111) substrate due to screening of dipoles parallel to the metal substrate.

The relative contributions of nuclear and electronic effects found here agree well with the experimental observation regarding Fano line shapes of IR absorption bands which may display different signs of the asymmetry parameter \cite{Braatz2012}. The main factor here is whether the dynamic dipole moments due to nuclear displacement and IDCT point in the same or opposite directions. Thus we can elucidate the interdependence of both mechanisms via a detailed analysis of the intensity data and derive the relative signs from ab initio computations.	

In conclusion, we have investigated electron-vibron coupling effects associated with vibrational excitations at metal-organic interfaces for the model system NTCDA on Ag(111).
Using infrared absorption spectroscopy and employing density functional theory based bonding and vibrational analysis we derived unequivocal evidence for the dominating role of interfacial dynamical charge transfer (IDCT) for dynamic dipole moments and associated infrared activities. Nuclear motion (out-of-plane bending) is found to be a secondary mechanism only.
The magnitude of IDCT for totally symmetric vibrational modes can be understood by pDOS shifts and derived from partial charge analysis. Contributions from nuclear motion can be determined reliably if IDCT is the leading term. This provides a simple, quantitative measure for the complex phenomenon. 
Our general approach will facilitate the analysis of other relevant metal-organic interfaces with similar matching of energy levels in the future.
On the basis of the thus obtained improved understanding, fine-tuning of the atomic and electronic structure for these types of interfaces becomes feasible.

\section{Methods}
\subsection{Computational details}
DFT computations at the GGA level (PBE-D3) \cite{PBE,GrimmeD3,GrimmeD3BJ} were carried out in a plane wave approach using the projector augmented wave method with an energy cutoff of $350\;\mathrm{eV}$ and a $\Gamma$-centered (3 3 1) Monkhorst-Pack k-space grid within the code VASP 5.2.12 \cite{VASP1996a,VASP1996b,PAW_Blochl,PAW_Kresse}. The adsorption structure of one NTCDA molecule on a four-layer Ag(111) slab (lattice parameter determined as \textit{a} = 4.073~\AA) was optimized in a $4\times4$ supercell. IR spectra for relaxed structures (without employing empirical scaling factors) were derived via finite-differences calculation of a partial Hessian matrix, displacing adsorbate atoms (a negligible influence of substrate atom displacement was found) and computing intensities based on the z-component of $\mu_{\mathrm{dyn}}$. This level of approximation was successfully used before for this system.\cite{Tonner2015_NTCDA_fullpaper}
IDCT has been deduced by displacing the equilibrium structure along the respective normal coordinates in positive ($+ \Delta$Q) and negative ($- \Delta$Q) direction. For the displaced structures, projected density of states (pDOS) were derived as well as atomic partial charges with the Natural Population Analysis scheme. \cite{Reed1985NPA,periodicNBO2012} Tests with the electron density-based Atoms In Molecules (AIM)\cite{Bader} scheme gave comparable results for $\Delta$q. 

\subsection{Experimental details}
The IRAS experiments were carried out in a UHV chamber at a base pressure of $\mathrm{5\times10^{-11}}$ mbar.
The Fourier-transform infrared spectrometer (FTIR) was a Bruker IFS 66v/S with evacuable optics.
The spectra were recorded using a liquid $\mathrm{N_{2}}$ - cooled Mercury-Cadmium-Telluride detector (spectral range $600-5000 \mathrm{cm}^{-1}$). In order to reduce the noise level, the displayed spectrum represents the sum of several consecutive measurements obtained at a sample temperature of 28 K and using polarized light; in total, 12500 scans at a spectral resolution of $2 \mathrm{cm}^{-1}$ have been co-added for the sample and the reference spectra, each.
The Ag(111) crystal was cleaned by $\mathrm{Ar^{+}}$ - sputtering (30 min at $\mathrm{U_{ion}= 700 eV}$,
$\mathrm{I_{ion}= 3 \mu A}$, and $\mathrm{T_{sample}= 373 K}$),  followed by annealing to 773 K for 5 min.
NTCDA was evaporated from a custom-made thermal evaporator at $\mathrm{T_{NTCDA}= 390 K}$
(deposition rate of 0.2 ML/min) controlled by a Pt1000 temperature sensor. During evaporation, the
background pressure typically increases by $\Delta p = \mathrm{1\times10^{-10}}$ mbar. 

\begin{acknowledgement}

This work was supported by the German Research Foundation (DFG) through collaborative research center `Structure and Dynamics of Internal Interfaces` (SFB~1083) and research training group `Functionalization of semiconductors` (GRK~1782). The authors thank the CSC-LOEWE Frankfurt and HLRS Stuttgart for computational resources.

\end{acknowledgement}

%
%
%


\begin{mcitethebibliography}{40}
\bibitem[Not()]{Note-1}
Dynamic dipole moments for adsorbate-only calculations were determined by
  PBE/def2-TZVPP computations with the code Gaussian09. $\mu_{\Delta\text{Q}}$
  were then derived by adding up the dipole moment changes from calculations of
  distortions in $\mathrm{+} \Delta\text{Q}$ and $\mathrm{-}\Delta\text{Q}$
  directions.\relax
\mciteBstWouldAddEndPunctfalse
\mciteSetBstMidEndSepPunct{\mcitedefaultmidpunct}
{}{\mcitedefaultseppunct}\relax
\EndOfBibitem
\bibitem[Not()]{Note-2}
Modes without significant $\mu_{nucl}$ were considered in the fit.\relax
\mciteBstWouldAddEndPunctfalse
\mciteSetBstMidEndSepPunct{\mcitedefaultmidpunct}
{}{\mcitedefaultseppunct}\relax
\EndOfBibitem
\end{mcitethebibliography}


\end{document}